\documentclass[aps,prd,superscriptaddress,nofootinbib,tighten,preprint]{revtex4-1}
\pdfoutput=1
\usepackage[utf8]{inputenc}
\usepackage[english]{babel}
\usepackage{amsmath}
\usepackage{subfigure}
\usepackage{hyperref}
\usepackage{amsfonts}
\usepackage{amssymb}
\usepackage{graphicx}
\usepackage{parskip}
\newcommand{\be}{\begin{equation}}
\newcommand{\ee}{\end{equation}}
\newcommand{\bea}{\begin{eqnarray}}
\newcommand{\eea}{\end{eqnarray}}
\newcommand{\ba}{\begin{align}}
\newcommand{\ea}{\end{align}}

\catcode`@=12

\begin{document}
\title{Inflation and Dark Matter in the Inert Doublet Model}

\author{Sandhya Choubey}
\email{sandhya@hri.res.in}
\affiliation{Harish-Chandra Research Institute, Chhatnag Road,
Jhunsi, Allahabad 211 019, India}
\affiliation{Homi Bhabha National Institute,
Training School Complex, Anushaktinagar, Mumbai - 400094, India}
\affiliation{Department of Physics, School of
Engineering Sciences, KTH Royal Institute of Technology, AlbaNova
University Center, 106 91 Stockholm, Sweden}
\author{Abhass Kumar}
\email{abhasskumar@hri.res.in}
\affiliation{Harish-Chandra Research Institute, Chhatnag Road,
Jhunsi, Allahabad 211 019, India}
\affiliation{Homi Bhabha National Institute,
Training School Complex, Anushaktinagar, Mumbai - 400094, India}
\begin{abstract}
We discuss inflation and dark matter in the inert doublet model coupled non-minimally to gravity where the inert doublet is the inflaton and the neutral scalar part of the doublet is the dark matter candidate. We calculate the various inflationary parameters like $n_s$, $r$ and $P_s$ and then proceed to the reheating phase where the inflaton decays into the Higgs and other gauge bosons which are non-relativistic owing to high effective masses. These bosons further decay or annihilate to give relativistic fermions which are finally responsible for reheating the universe. At the end of the reheating phase, the inert doublet which was the inflaton enters into thermal equilibrium with the rest of the plasma and its neutral component later freezes out as cold dark matter with a mass of about 2 TeV.
\end{abstract}
\maketitle

\section{Introduction}

The standard model of particle physics has been very successful with highly accurate predictions. However, it still has no answer for various problems like dark matter and inflation. Both inflation and dark matter have been established very firmly through various observations particularly of the cosmic microwave background (CMB) radiation. Inflation \cite{Guth:1980zm} has long been the most successful theory to answer cosmological problems like the horizon problem and homogeneity. The most popular inflationary models are those that have an extra scalar particle which acts as the inflaton. Recent experiments like Planck \cite{Ade:2015lrj} and WMAP7 \cite{Komatsu:2010fb} have placed bounds with high accuracy on inflationary parameters like the spectral index, the tensor to scalar ratio and the scalar power spectrum. There have been a variety of inflation models over the years. The Higgs inflation \cite{Bezrukov2008703, Bezrukov2011} models are the most simple in the sense that they do not involve any extra field and have just one more parameter $\xi$ through which the field couples to gravity but they come with their share of problems. The quartic coupling $\lambda$ of Higgs field at high energy scales ($\gtrsim10^{10}$ GeV) becomes negative. This can cause problems with the stablility of the vacuum \cite{SHER1989273}. Another problem comes in the form of non-unitarity. The scalar power spectrum bounds require $\xi\sim 10^4$ \cite{1475-7516-2010-04-015} which breaks unitarity at scales around $m_{Pl}/\xi\approx 10^{13}$ GeV  \cite{Burgess2010}. To avoid running into problems in a Higgs inflation model, often an extra scalar stabilizing field is added and such scenarios are called $s$-inflation. These models have an extra gauge singlet scalar particle that acts as the inflaton while the Higgs field acts as a portal to the standard model to reheat the universe. There can be variations in this model and in \cite{Lerner} distinctions between the variations is studied. The inflationary potential is usually taken to be either a chaotic one or a Starobinsky one. Chaotic inflation \cite{LINDE1983177} models include power law potentials like $m^2\phi^2+\lambda\phi^4$. These were the first type of potentials used to study inflation. On the other hand Starobinsky models have exponential potentials. We will discuss more about them later. A good review for inflationary cosmology in the light of data can be had in \cite{Linde:2014nna}.

Dark matter has been studied extensively over the years. Thanks to the many experiments and observations, we now have a good estimate for dark matter distribution in and around our galaxy and in the universe at large. Planck results \cite{Ade:2015xua} together with other astronomical observations have put down the abundance of dark matter in the universe to $\Omega_{dm} h^2\simeq 0.12$. The most commonly studied dark matter scenarios are the so called Weakly-Interacting Massive Particles (WIMP). In recent years however, as dark matter detection experiments have become better and colliders like LHC are probing higher energies, the absence of any new particle at the weak scale has put the WIMP scenario in a fix and people have started looking at other options like axions, feebly interacting massive particles (FIMP) and strongly interacting massive particles (SIMP) \cite{Hall:2009bx, Yaguna2011, Baer20151, PhysRevLett.113.171301} among others. One of the simplest models of dark matter -- the scalar singlet dark matter model is still being sustained and there have been updates to it \cite{PhysRevD.88.055025}; see also \cite{Queiroz:2014yna}. 

In more recent works, people have started to look for scenarios where both inflation and dark matter can be explained by the same field. Gauge singlet scalar models in the $s$-inflation scenario is a case in point. In this paper, we have combined inflation and dark matter in the inert doublet model coupled non-minimally to gravity. Such a unification was first shown to be possible in \cite{Liddle} in string theory landscape. In \cite{1475-7516-2015-11-015, Lerner:2009xg, PhysRevD.93.123513} a gauge singlet scalar is used as inflation and later after freeze out as the dark matter candidate. \cite{Tenkanen} has a situation similar to $s$-inflation where the inflaton is very light and interacts very feebly to become FIMP dark matter later. Inflation and dark matter in two Higgs doublet models was studied in \cite{Gong}. A scalar WIMP dark matter candidate with non-minimal coupling to gravity acting as the inflaton was studied in \cite{Okada:2010jd}.

The motivation for using inert doublet model in our case is the fact that pure Higgs inflation is problematic and yet it is the only scalar field present in the standard model. Another scalar doublet similar to Higgs doublet but stabilized by an extra $\mathbb{Z}_2$ symmetry such that it does not interact with leptons and quarks via Yukawa couplings can present a viable candidate for both inflation and dark matter. The components of the inert doublet can all act as inflaton via a particular field redefinition. At the same time, its neutral scalar component can later become the dark matter candidate. The inert doublet through its interactions with the vector gauge bosons and Higgs can also reheat the universe at the end of inflation to ensure that the universe gets populated by standard model particles. Another motivation for using this model is that it is similar to $s$-inflation models in that the potential turns out to be of the Starobinsky kind which gives some of the best fit to inflationary parameters like the spectral index. We will also look at the reheating phase in some detail. Inflaton during reheating behaves as non-relativistic matter and decays via gauge and Higgs bosons to relativistic particles. We will look at the interactions happening during reheating and later when we discuss dark matter, we will point out the changes that take place in the interactions of the inert doublet compared to the reheating phase. The electroweak (EW) symmetry breaking will play a role in determining the type of interactions that the inert doublet undergoes.

This paper is organized in the following manner. We describe the model in the next section. In section 3, we study inflation and find the value of the various inflationary parameters like the slow roll parameters, the spectral index and the tensor to scalar ratio. In section 4 we study reheating which progresses by the decay of the inflaton into non-relativistic vector and Higgs bosons which further annihilate into relativistic fermions. In this section, we calculate the energy density stored in the relativistic particles and find some bounds on some model parameters. The inert doublet as a cold dark matter candidate is taken up in section 5 where we fix some parameter values like the mass of dark matter through relic density calculations. We end in section 6 with conclusions.

\section{The model}
We will use the inert doublet model coupled non-minimally to gravity where there is an extra doublet $\Phi_2$ apart from the Higgs doublet $\Phi_1$. The extra doublet is inert in the sense that it does not have any Yukawa like couplings because of an inherent $\mathbb{Z}_2$ symmetry under which this doublet is odd ($\Phi_2\rightarrow -\Phi_2$) while the Higgs and other standard model particles are even ($\Phi_1,\psi\rightarrow \Phi_1,\psi$, where $\psi$ stands for SM particles other than Higgs). The action of this model is:

\begin{eqnarray}\label{action}
S=\int d^4x \sqrt{-g}\left[-\frac{1}{2}M_{Pl}^2 R-D_{\mu}\Phi_1 D^{\mu}\Phi_1^\dagger-D_{\mu}\Phi_2 D^\mu \Phi_2^\dagger-V(\Phi_1,\Phi_2)-\xi_1\Phi_1^2 R-\xi_2\Phi_2^2 R\right],
\end{eqnarray}
where $D$ stands for the covariant derivative containing couplings with the gauge bosons. During inflation, there are no fields other than the inflaton so that the covariant derivative will reduce to the normal derivative $D_\mu\rightarrow \partial_\mu$ The minus sign in the kinetic terms is in keeping with the metric convention of $(-,+,+,+)$. $M_{Pl}$ is the reduced Planck mass, $R$ is the Ricci scalar and $\xi_1$ and $\xi_2$ are dimensionless couplings of the doublets to gravity. The motivation behind including these couplings is that quantum effects invariably give rise to such couplings at Planck scales \cite{Birrell}.

The potential is:

\begin{eqnarray}\label{OriPot}
V&=&m_1^2 |\Phi_1|^2 + m_2^2|\Phi_2|^2 + \lambda_1(|\Phi_1|^2)^2 + \lambda_2(|\Phi_2|^2)^2\nonumber\\
&&\;\;\;\;+\lambda_3|\Phi_1|^2|\Phi_2|^2+\lambda_4(\Phi_1^\dagger \Phi_2)(\Phi_2^\dagger \Phi_1) + \frac{1}{2}\lambda_5[(\Phi_1^\dagger \Phi_2)^2+c.c.].
\end{eqnarray}
The two doublets have the components:
\begin{equation}
\Phi_1=\frac{1}{\sqrt{2}}\left(\begin{array}{c}\chi\\h\end{array}\right)\;\; \mathrm{and}\;\; \Phi_2=\frac{1}{\sqrt{2}}\left(\begin{array}{c}q\\ x\,e^{\mathtt{i}\theta}\end{array}\right).
\end{equation} 

Note that there is no non-zero vacuum expectation value of the Higgs field as the electroweak symmetry is intact at inflationary scales. We want the inert doublet to be the inflaton. This is ensured if $\frac{\lambda_2}{\xi_2^2}\ll \frac{\lambda_1}{\xi_1^2}$. A choice where $\lambda_1$ and $\xi_1$ are of the same order while $\lambda_2\sim 1\ll\xi_2$ automatically satisfies this condition

\section{Inflation}
The action in Eq. (\ref{action}) is written in the physical or the Jordan frame \cite{0264-9381-14-12-010, PhysRevD.81.084044} and has terms where the scalars $\Phi_{1,2}$ couple quadratically to gravity. This makes it difficult to derive meaningful results from the usual processes of quantum field theory. We need to make some transformations where we can get rid of such coupled terms. This can be done by a conformal transformation to the so called Einstein frame. Einstein frame is useful as in this frame the action looks like a regular field theory action with no explicit couplings to gravity. Results for physical observables remain the same independent of the frame chosen. After the end of inflation, the transformation parameter becomes almost 1, making the two frames equivalent. Following \cite{PhysRevD.81.084044}, 
we make the following conformal transformation on the metric and the fields to get the action in the Einstein frame:
Defining $\phi=\{\chi,h,q,x,\theta\}$
\begin{eqnarray}
S=\int d^4x \sqrt{-\tilde g}\left[-\frac{1}{2}M_{Pl}^2\tilde R-\frac{1}{2}G_{ij}\tilde{g}^{\mu\nu}\partial_\mu \phi_i \partial_\nu\phi_j-\tilde{V}(h,q,x,\theta)\right],
\end{eqnarray}
where:
\begin{eqnarray}
\tilde{g}_{\mu\nu}&=&\Omega^2 g_{\mu\nu}, \\
\Omega^2 &=&1+\frac{\xi_1}{M_{Pl}^2}(\chi^2+h^2)+\frac{\xi_2}{M_{Pl}^2}(q^2+x^2),\\
G_{ij}&=& \frac{1}{\Omega^2}\delta_{ij}+\frac{3}{2}\frac{M_{Pl}^2}{\Omega^4}\frac{\partial\,\Omega^2}{\partial\,\phi_i}\frac{\partial\,\Omega^2}{\partial\,\phi_j},  \\
\tilde{V}&=&\frac{V}{\Omega^4}. \label{PotDef}
\end{eqnarray}

Let us look at the kinetic terms. First, we expand the pre-factor $G$ in a matrix form:

\begin{eqnarray}
G=\left[ \begin{array}{ccccc} \frac{\Omega^2+6\xi_1^2 \chi^2/M_{Pl}^2}{\Omega^4} & 6\frac{\xi_1^2}{M_{Pl}^2\Omega^2}\chi\,h & \frac{6\xi_1 \xi_2}{M^2\Omega^4} \chi\,q & \frac{6\xi_1 \xi_2}{M^2\Omega^4} \chi\,x & 0 \\&&&& \\ 6\frac{\xi_1^2}{M_{Pl}^2\Omega^2}\chi\,h & \frac{\Omega^2+6\xi_1^2 h^2/M_{Pl}^2}{\Omega^4} & \frac{6\xi_1 \xi_2}{M^2\Omega^4} hq & \frac{6\xi_1 \xi_2}{M^2\Omega^4} hx & 0 \\ & & & & \\ 
\frac{6\xi_1 \xi_2}{M^2\Omega^4} \chi\,q & \frac{6\xi_1 \xi_2}{M^2\Omega^4} hq & \frac{\Omega^2+6\xi_2^2 q^2/M_{Pl}^2}{\Omega^4} & \frac{6 \xi_2^2}{M^2\Omega^4} qx & 0 \\ &&&& \\
\frac{6\xi_1 \xi_2}{M^2\Omega^4} \chi\,x & \frac{6\xi_1 \xi_2}{M^2\Omega^4} hx & \frac{6\xi_2^2}{M^2\Omega^4} qx & \frac{\Omega^2+6\xi_2^2 x^2/M_{Pl}^2}{\Omega^4} & 0 \\ &&&&\\0 & 0 & 0 & 0 & \frac{x^2}{\Omega^2} \end{array} \right].
\end{eqnarray}

The above $G$ gives mixed kinetic terms. All these fields are always present in the lagrangian but during inflation, fields other than the inert doublet components give no contribution. $\Omega^2$ can also be simplified to exclude the $\frac{\xi_1}{M_{Pl}} (\chi^2+h^2)$ term. This allows us to simplify the matrix $G$ as:
\begin{eqnarray}\label{newG}
G=\left[ \begin{array}{ccccc} \frac{1}{\Omega^2} & 0 & 0 & 0 & 0 \\&&&&\\ 0 &  \frac{1}{\Omega^2} & 0 & 0 & 0 \\ & & & & \\ 
0 & 0 & \frac{\Omega^2+6\xi_2^2 q^2/M_{Pl}^2}{\Omega^4} & \frac{6 \xi_2^2}{M^2\Omega^4} qx & 0 \\ &&&& \\
0 & 0 & \frac{6\xi_2^2}{M^2\Omega^4} qx & \frac{\Omega^2+6\xi_2^2 x^2/M_{Pl}^2}{\Omega^4} & 0 \\ &&&&\\0 & 0 & 0 & 0 & \frac{x^2}{\Omega^2} \end{array} \right].
\end{eqnarray}

A further simplification to a completely diagonal kinetic form can be obtained by rearranging the fields as follows:
\begin{eqnarray}
A&=&\sqrt{\frac{3}{2}}M_{Pl}\log\left(\Omega^2\right),\label{defA}\\
B&=M_{Pl}\,&\frac{x}{q}.
\end{eqnarray}

Substituting this redefinition of fields into the kinetic part, we get a diagonal kinetic term which is:
\begin{eqnarray}\label{kinet}
\frac{1}{2\Omega^2}\left((\partial_\mu\chi)^2+(\partial_\mu h)^2\right)+\left[\frac{1}{2}+\frac{1}{12\xi_2 F(A)}\right](\partial_\mu A)^2+\nonumber\\\left[\frac{F(A)}{2\xi_2(1+B^2/M_{Pl}^2)^2}\right](\partial_\mu B)^2+\left[\frac{F(A)\,B^2}{2\xi_2(1+B^2/M_{Pl}^2)}\right](\partial_\mu\theta)^2,
\end{eqnarray}
where $F(A)=1-\exp\left(-\sqrt{\frac{2}{3}}\frac{A}{M}\right)$.

\begin{figure}
\centering
\includegraphics[scale=1]{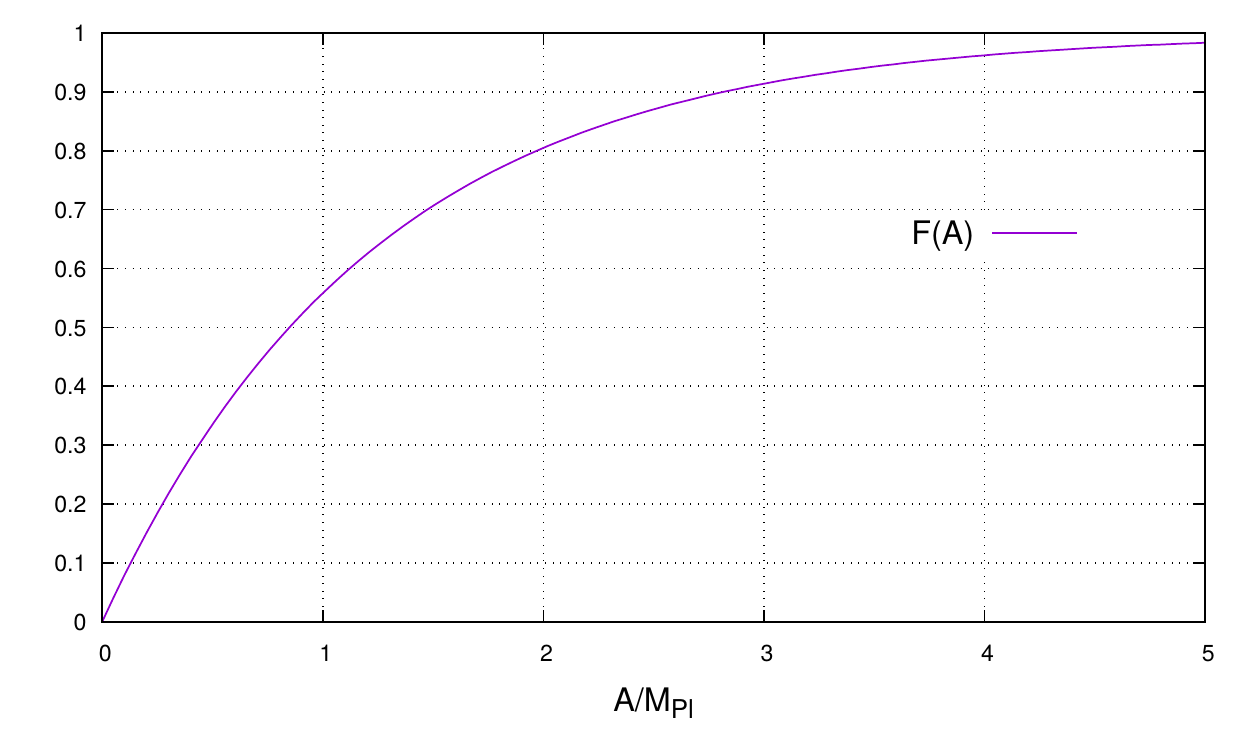}
\caption{The graph of $F(A)$}
\label{Fig0}
\end{figure}

Eq. (\ref{kinet}) is still apparently not canonical. However, at the scales relevant for inflation $F(A)$ is of the order of 1 and the change in $F(A)$ while $A$ drops from values many times larger than $M_{Pl}$ to  $M_{Pl}$ is very small. This can be seen in Fig. \ref{Fig0}. With large $\xi_2$ this means that the coefficient of $(\partial_\mu A)^2\approx \frac{1}{2}$ and the other fields can have a constant rescaling which makes the kinetic term canonical.

All such terms from the Einstein frame potential in Eq. \ref{PotDef} which are not quartic in $q$ and $x$ can be neglected owing to the largeness of these two fields. The only relevant term that remains is the  quartic term $\frac{1}{4}\lambda_2\left(q^2+x^2\right)^2$ which using the redefined fields becomes:
\begin{eqnarray}\label{potential}
V_e\approx \frac{\lambda_2 M_{Pl}^4}{4\xi_2^2}\left[1-\exp\left(-\sqrt{\frac{2}{3}}\frac{A}{M_{Pl}}\right)\right]^2.
\end{eqnarray}

\begin{figure}
\centering
\includegraphics[scale=1]{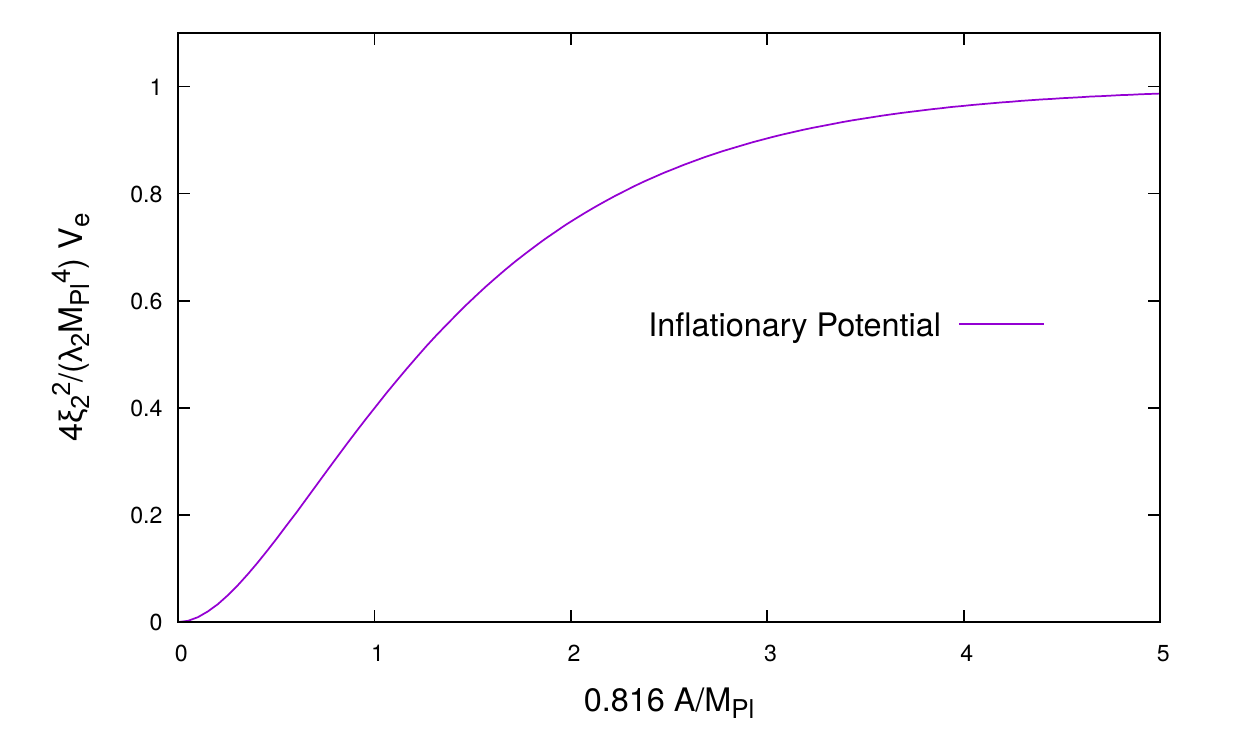}
\caption{\textit{The slowly rolling inflationary potential}}
\label{Fig1}
\end{figure}

The potential in Eq. (\ref{potential}) belongs to a class of potentials called the Starobinsky potentials \cite{Starobinsky:1979ty}; see also \cite{Calmet:2016fsr}. In Fig. \ref{Fig1} we show the inflationary potential vs. the field where it can be seen that the potential is almost flat at high field values ensuring slow roll. The slow roll parameters $\epsilon$ and $\eta$ with this potential are:


\begin{eqnarray}
\epsilon = \frac{1}{2}M_{Pl}^2\left(\frac{1}{V_e}\frac{dV_e}{dA}\right)^2&=&\frac{4}{3}\left[-1+\exp\left(\sqrt{\frac{2}{3}}\frac{A}{M_{Pl}}\right)\right]^{-2}\label{epsilon}\,,\\
\eta = M_{Pl}^2\frac{1}{V_e}\frac{d^2\,V_e}{dA^2}&=&\frac{4}{3}\frac{\left[2-\exp\left(\sqrt{\frac{2}{3}}\frac{A}{M_{Pl}}\right)\right]}{\left[-1+\exp\left(\sqrt{\frac{2}{3}}\frac{A}{M_{Pl}}\right)\right]^2}\,.
\end{eqnarray}

For field values $A\gg M_{Pl}$, both $\epsilon,\eta \ll 1$ and thus slow roll is satisfied. Inflation ends when $\epsilon \simeq 1$.

We would now like to get estimates for the values of $A$ at the beginning and end of inflation which will be needed to get the power spectrum. This can be done by looking at the number of times the universe expanded by $e$ times its own size, also called the number of e-folds $N$. It is obtained as follows:
\begin{eqnarray}
N&=&\frac{1}{M_{Pl}^2}\int_{A_{end}}^{A_{ini}} \frac{V_e}{V_e'}\,dA\nonumber\\
&=&\frac{3}{4}\left[\exp\left(\sqrt{\frac{2}{3}}\frac{A_{ini}}{M_{Pl}}\right)-\exp\left(\sqrt{\frac{2}{3}}\frac{A_{end}}{M_{Pl}}\right) -   \sqrt{\frac{2}{3}}\frac{A_{ini}}{M_{Pl}}+\sqrt{\frac{2}{3}}\frac{A_{end}}{M_{Pl}}   \right] \,,\label{efolds}
\end{eqnarray}
where $V_e'=\frac{dV_e}{dA}$, $A_{ini}$ is the value of $A$ at the beginning of inflation and $A_{end}$ is the value of $A$ at the end of the inflation.

To get $A_{end}$, we make use of the fact that slow roll inflation ends when $\epsilon\simeq 1$ in Eq. (\ref{epsilon}), which gives:
\begin{eqnarray}
\exp\left(\sqrt{\frac{2}{3}}\frac{A_{end}}{M_{Pl}}\right)&\simeq & 2.15, \label{eps1} \\
\sqrt{\frac{2}{3}}\,\frac{A_{end}}{M_{Pl}}&\simeq & 0.77.
\end{eqnarray}

Using Eq. (\ref{eps1}) in Eq. (\ref{efolds}) for $N=60$ \footnote{In principle $N$ could be any number greater than around $50$ to solve flatness and horizon problems. $60$ e-folds solves the baryon asymmetry problem if inflationary energy scales are $O[10^{16}]\,\textrm{GeV}$ \cite{PhysRevD.78.123501}. Lower inflationary energy scales would need more e-folds and vice versa. However, the number of e-folds cannot be much larger than $60$.} we get
\begin{eqnarray}
\frac{3}{4}\left[\exp\left(\sqrt{\frac{2}{3}}\frac{A_{ini}}{M_{Pl}}\right)-\sqrt{\frac{2}{3}}\frac{A_{ini}}{M_{Pl}}-1.387  \right]&=&60, \\
\Rightarrow \sqrt{\frac{2}{3}}\frac{A_{ini}}{M_{Pl}}&\approx & 4.45.
\end{eqnarray}

Looking at Fig. \ref{Fig1}, we see that field values are consistent with slow-roll and its end.

\begin{figure}[h!]
\centering
  \includegraphics[scale=1]{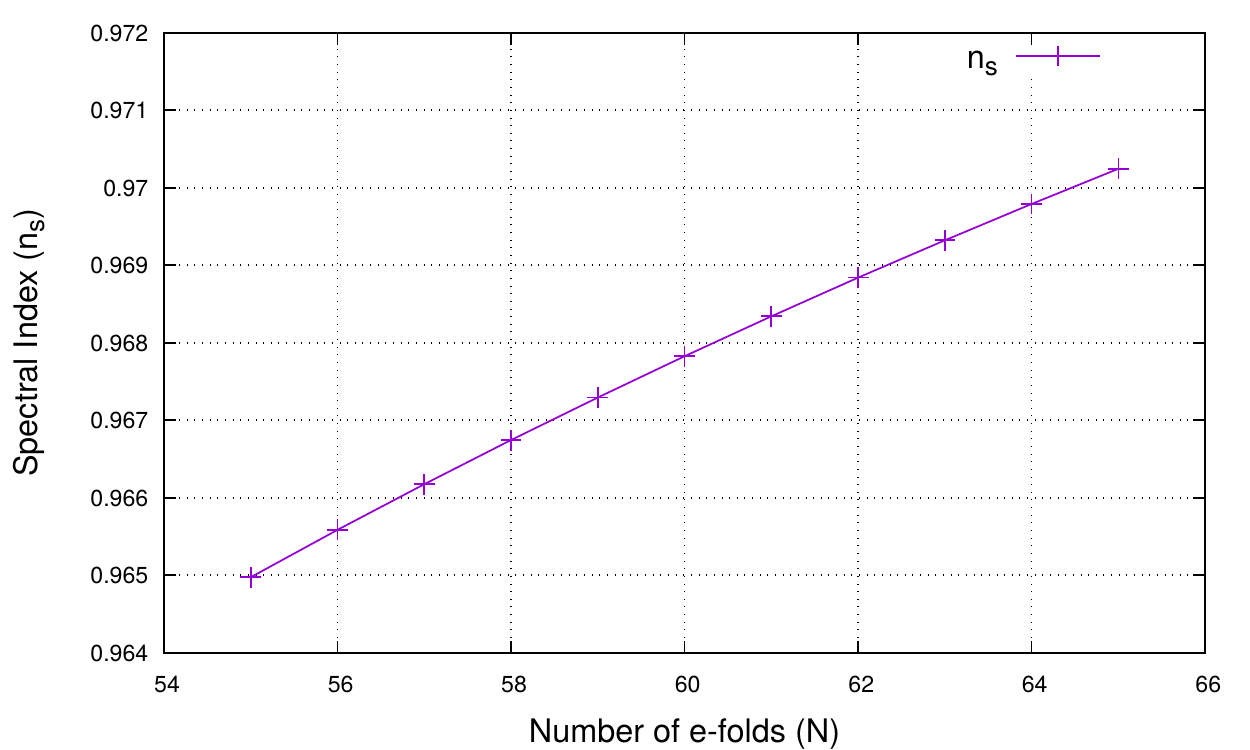}
  \caption{\textit{The spectral index as a function of N}}
  \label{Fig2}
\end{figure}

\begin{figure}[h!]
  \centering
  \includegraphics[scale=1]{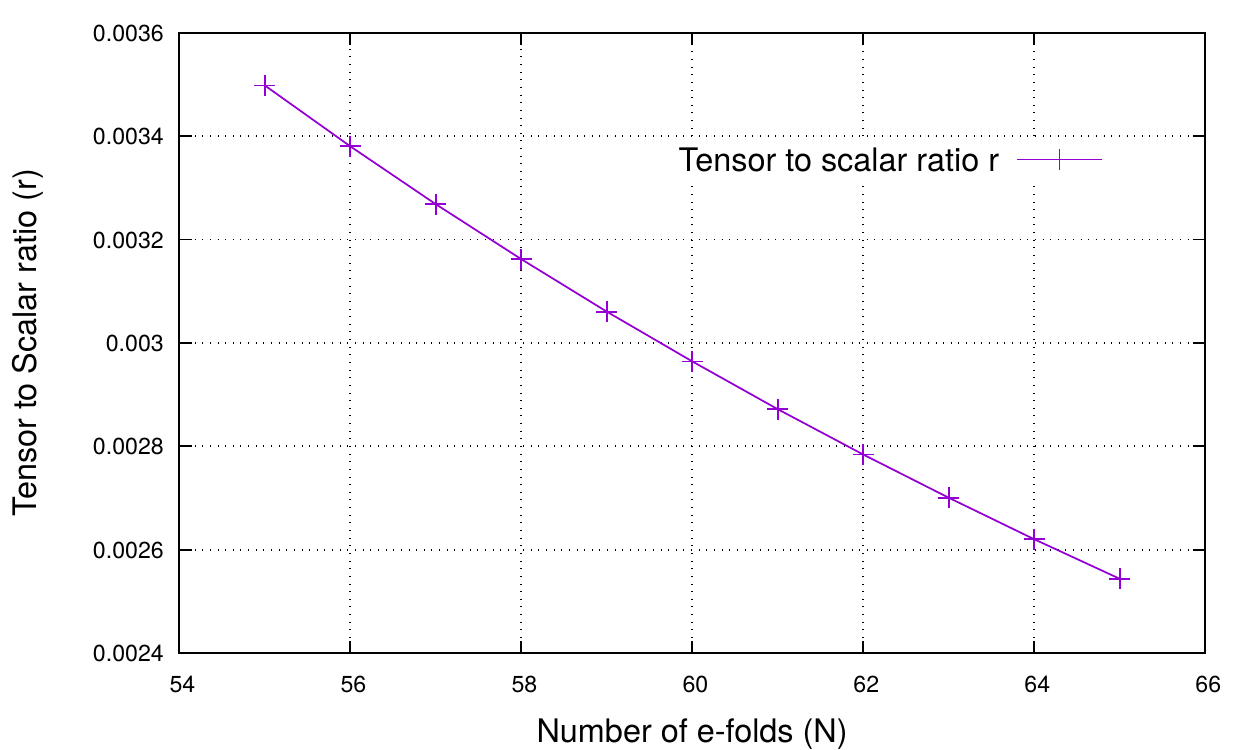}
  \caption{\textit{Tensor to scalar ratio as a function of N}}
  \label{Fig3}
\end{figure}

With $N$ fixed at 60 and the field value at the start of inflation fixed, we can get the scalar power spectrum ($P_S$), the tensor to scalar ratio ($r$) and the spectral index ($n_s$) as follows:

\begin{eqnarray}
P_s =\frac{1}{12\,\pi^2}\frac{V_e^3}{M_{Pl}^6\,V_e^{'2}}&=&5.57\times\frac{\lambda_2}{\xi_2^2},\\
\nonumber\\
r =16\,\epsilon &=&0.0029, \\
n_s =1-6\epsilon+2\eta &=& 0.9678,
\end{eqnarray}

where $V_e'$ is the derivative of $V_e$ with respect to $A$ and both $V_e$ and $V_e'$ are calculated at the $A_{ini}$. The values of $r$ and $n_s$ are well within the Plank bounds \cite{Ade:2015lrj} of $n_s=0.9677\pm 0.0060 $ at $1\sigma$ level and $r<0.11$ at $95\%$ confidence level. Since there is no reason for $N$ to be precisely $60$, we look at the inflationary parameters over a range of $N$ from 55 to 65 (see Fig. (\ref{Fig2}) and (\ref{Fig3})). We see that in the entire region of $N$, the spectral index and the tensor to scalar ratio lie within Planck bounds.

We can use WMAP7 constraint for $P_s$ \cite{Komatsu:2010fb} to relate $\lambda_2$ and $\xi_2$ which will be needed later for energy density calculations.
\begin{eqnarray}
P_s=(2.430\pm 0.091)\times 10^{-9}&=&5.57\frac{\lambda_2}{\xi_2^2},\nonumber\\
\Rightarrow \xi_2&\approx& 4.79\times 10^4 \,\lambda_2^{1/2}.
\end{eqnarray}

\subsection{A note on isocurvature fluctuations}
Having multiple scalar fields can give rise to multi-field effects which can cause significant iso-curvature fluctuations. The presence of isocurvature fluctuations has been studied in detail in \cite{Kaiser:2013sna, Schutz:2013fua, Greenwood:2012aj}. Following them, we expand the fields to first order $\phi^i=\varphi^i(t)+\delta\phi^i(x^\mu)$ and define
\begin{eqnarray}
\dot{\sigma}^2&=&G_{ij}\dot{\varphi^i}\dot{\varphi^j} \\
\hat{\sigma}^i&=&\frac{\dot{\varphi^i}}{\dot{\sigma}} \\
\hat{s}^i&=&\frac{\varepsilon^i}{\varepsilon}
\end{eqnarray}
where $\varepsilon$ is the turn-rate vector in the field space: $\varepsilon^i=\dot{\hat{\sigma}}^i+\Gamma^i_{jk}\hat{\sigma}^j\,\dot{\varphi}^k$ with $\Gamma^i_{jk}$ being the connection in the field space for the field space metric $G_{ij}$. We also define the mass-squared matrix for the gauge invariant linearized perturbations \cite{Kaiser:2013sna}:
\begin{equation}
M^i_j=G^{ij}\mathcal{D}_j\mathcal{D}_k\,Ve-\mathcal{R}^i_{klj}\dot{\varphi}^k\dot{\varphi}^l
\end{equation}
where $\mathcal{D}_i$ is the covariant derivative in the field space wrt field $\varphi^i$ and $\mathcal{R}^i_{klj}$ is the Riemann tensor in field space. These together are used to get a parameter $\eta_{ss}$ which is used to calculate the mass-square of the isocurvature fluctuations $\mu_s^2$ as follows:
\begin{eqnarray}
\eta_{ss}&=&\frac{\hat{s}_i\hat{s}^j\,M^i_j}{Ve}\,M_{Pl}^2\\
\mu_s^2 &=&3\,H^2\,(\eta_{s}+\frac{\varepsilon ^2}{H^2})
\end{eqnarray}

Since $\frac{\lambda_2}{\xi_2^2}\ll\frac{\lambda_1}{\xi_1^2}$, inflation occurs along the $\chi\sim h\sim 0$ direction, thereby making $\hat{s}^1$ and $\hat{s}^2$ zero. We are left with remaining two scalars $q$ and $x$ which have symmetric couplings $\lambda_2$ and $\xi_2$. For such a case, $\eta_{ss}\ll 1$ $(\sim O(10^{-6}))$ which means \cite{Greenwood:2012aj} $\mu_s^2/H^2\simeq 0$ giving a very suppressed isocurvature fraction of $\beta_{iso}\sim O(10^{-5})$. The results are hence consistent with Planck data  \cite{Ade:2015lrj}. 

\section{Reheating}
At the end of inflation, the universe is too dilute for anything to be present. Unless the universe is somehow repopulated by particles, it remains empty. It is at this juncture that the energy density till now stored in the inflaton starts to disperse as the inflaton particles annihilate or decay into other particles including those of the standard model. This phase of the universe after inflation where inflaton annihilates into other relativistic particles is called reheating \cite{Linde:1981mu}. If the inflaton decays or annihilates into bosons, parametric resonance production of bosons triggers efficient reheating \cite{Dolgov:1989us, Traschen:1990sw} (see also \cite{PhysRevLett.73.3195, Shtanov:1994ce}) and at the end of it, the universe becomes radiation dominated.

The conformal transformation and the redefinition of fields done in the previous section allows us to identify two distinct regions \cite{Bezrukov} marked by $A_{cr}=\sqrt{\frac{2}{3}}\frac{M_{Pl}}{\xi_2}$:  

\begin{equation}
A\approx \Bigg{\lbrace} \begin{array}{c} (q^2+x^2)^{1/2}\;\;\;\;\textrm{for }A<A_{cr}, \\ \sqrt{\frac{3}{2}}\,M_{Pl}\,\log\left(\Omega^2\right)\;\;\;\;\textrm{for }A>A_{cr}.
\end{array}
\end{equation}

Inflation occurs in the second region where $A>A_{cr}$ which can also be written as $(q^2+x^2)^{1/2}>A_{cr}$. 
Much below $M_{Pl}$, the inflationary potential in Eq. (\ref{potential}) can be approximated by a quadratic potential well:

\begin{eqnarray}
V_e&=&\frac{\lambda_2\,M_{Pl}^4}{4\,\xi_2^2}\left[1-\exp\left(-\sqrt{\frac{2}{3}}\frac{A}{M_{Pl}}\right)\right]^{2}, \nonumber\\
&\simeq&\frac{\lambda_2M_{Pl}^2}{6\xi_2^2}A^2, \\
V_e&=&\frac{1}{2}\omega^2\,A^2,\;\;\textrm{where }\omega^2=\frac{\lambda_2\,M_{Pl}^2}{3\xi_2^2}, \label{PotRt}
\end{eqnarray}

This is a simple harmonic potential in which the inflaton oscillates rapidly with frequency $\omega$. This makes the oscillations coherent, the phase being the same at all points in space. Since the potential is a simple harmonic one near the minimum, the average energy density obeys the relation $\bar{\rho}_A=\langle\dot{A}^2\rangle$ and thus obeys the equation $\dot{\bar{\rho}}_A+3H\bar{\rho}_A=0$ which yields a $1/a^3$ evolution for the average energy density. This means that during this period the inflaton behaves as non-relativistic matter. A matter dominated universe has the following characteristics with respect to the scale factor and Hubble's constant:
\begin{eqnarray}
a&\propto &t^{2/3},\\
H(t)=\frac{\dot a(t)}{a(t)}&=&\frac{2}{3t}.\label{Hast}
\end{eqnarray}
The equation of motion for $A$ during this phase is 
\begin{equation}
\ddot{A}+3H\dot{A}+\frac{dV_e}{dA}=0,
\end{equation}
which gives on solving for $\omega\gg H$:
\begin{equation}
A=A_0(t)\cos (\omega t),
\end{equation}
where
\begin{equation}
A_0(t)=2\sqrt{2}\frac{\xi_2}{\sqrt{\lambda_2}}\frac{1}{t}.
\end{equation}

The quadratic phase ends when the amplitude of the oscillations $A_0$ crosses $A_{cr}$ which gives us the crossing time as $t_{cr}=\frac{2\xi_2}{\omega}$. In \cite{Linde:1981mu} it was shown that reheating occurs when the field oscillates in a quadratic potential well. Therefore in the present scenario, reheating starts when the potential gets approximated by Eq. (\ref{PotRt}) and ends when the amplitude $A_0$ crosses $A_{cr}$ at time $t_{cr}$.

\subsection{Decay of the inflaton}
The inert doublet can decay into the $W$ and $Z$ bosons through the kinetic coupling terms and into the Higgs boson through the potential in the Lagrangian. The resultant particles don't have a physical mass at this time but an effective mass arising due to the inflaton oscillations. If the oscillation frequency $\omega$ is much larger than the expansion rate $H$, the amplitude can be taken to be constant over one oscillation period. This allows us to write down effective mass terms for the vector and scalar bosons. When $A\ll \sqrt{\frac{3}{2}}\,M_{Pl}$ but still above $A_{cr}$, $(q^2+x^2)\ll \frac{M_{Pl}^2}{\xi_2}$. Using this we can expand the log term in the definition of $A$ in terms of $q$ and $x$ to get:
\begin{equation}\label{qxapprox}
q^2+x^2\simeq \sqrt{\frac{2}{3}}\frac{M_{Pl}}{\xi_2} A.
\end{equation}

The coupling of the inert doublet to $W$ bosons is $\frac{g^2}{4}(q^2+x^2)W^2$ which in terms of $A$ is $\frac{g^2}{4\sqrt{6}}\frac{M_{Pl}}{\xi_2} A\, W^2$. This gives an effective mass for $W$ bosons to be:
\begin{equation}\label{mW}
m_W^2=\frac{g^2}{2\sqrt{6}}\frac{M_{Pl}}{\xi_2}|A|.
\end{equation}
The other vector boson effective masses can be related by the Weinberg angle.

The coupling to Higgs is through $\lambda_{3,4,5}$. This gives us an effective mass term for the Higgs boson:
\begin{equation}\label{mh}
m_h^2=\frac{1}{\sqrt{6}}\left(\lambda_3+\frac{\lambda_4}{2}\right)\frac{M_{Pl}}{\xi_2}|A|.
\end{equation}
In writing the Higgs effective mass, we have taken equal contributions of $q$ and $x$ in $A$.

Note that Eq. (\ref{qxapprox}) is not an equation for vacuum expectation value as it is not calculated at the minimum of the potential. It just describes the transformation between $q$ and $x$ on one hand and $A$ on the other in a particular regime mentioned above the equation that follows from Eq. (\ref{defA}). The masses in Eqs. (\ref{mW}) and (\ref{mh}) are therefore not usual masses obtained from spontaneous symmetry breaking but are just effective masses coming out of their interactions with the inflaton fields when written in terms of the transformed field $A$.

The weak coupling $g$ is large which makes the vector bosons non-relativistic. They will decay and annihilate to other relativistic fermions to reheat the universe. If either of $\lambda_3$ and $\lambda_4$ is large, the produced Higgs too will be non-relativistic and it will decay into fermions through Yukawa interactions which will add to the relativistic energy density. The inert doublet gives a cold dark matter candidate which means the combination of its couplings to Higgs becomes of the order of $1$ \cite{Gustafsson:2010zz}. We choose a case where $\lambda_3\approx 1$ is the dominant coupling when compared to $\lambda_4$ and $\lambda_5$ which are taken to be very small just for the sake of brevity. This enables us to remove the $\lambda_4$ term from the effective mass of Higgs in Eq. (\ref{mh}).

At low number densities of the produced $W$ and Higgs bosons ($n_W$ and $n_h$ respectively) their decay to fermions is the dominant channel to produce relativistic  particles. If the number density of the bosons becomes large, their production rate will become exponential due to parametric resonancne. During the resonance phase, the $W$ bosons will mostly annihilate to produce fermions. Their decays will become sub-dominant channels of fermion production. Higgs on the other hand can only produce fermions through decays. Following \cite{Bezrukov} (see also \cite{GarciaBellido:2008ab, Repond:2016sol}), the production of $W$ bosons in the linear and resonance regions is:
\begin{equation}\label{Wprod}
\frac{d(n_W a^3)}{dt}=\left\{\begin{array}{c} \frac{P}{2\pi^3}\omega K_1^3 a^3,\;\;\;\; \textrm{(linear)}, \\ \\ 2\,a^3\,\omega Q\,n_W, \;\;\;\;\textrm{(resonance)},\end{array}\right.
\end{equation}
where $P$ and $Q$ are numerical factors with $P\approx 0.0455$ and $Q\approx 0.045$ and $\alpha_W=\frac{g^2}{4\pi}$ is the weak coupling constant.

Making the corresponding changes for the production of Higgs, we have:
\begin{equation}\label{hprod}
\frac{d(n_ha^3)}{dt}=\left\{\begin{array}{c} \frac{P}{2\pi^3}\omega K_2^3 a^3,\;\;\;\; \textrm{(linear)}, \\ \\ 2 a^3 \omega Q n_h.\;\;\;\; \textrm{(resonance)}.

\end{array}\right.
\end{equation}

$K_1$ and $K_2$ have dimensions of energy and are dependent on the respective mass terms with:
\begin{eqnarray}
K_1&=&\left[\frac{g^2 M_{Pl}^2}{6\xi_2^2}\sqrt{\frac{\lambda_2}{2}}A_0(t_i)\right]^{1/3}, \\
K_2 &=& \left[\frac{\lambda_3\,M_{Pl}^2}{3\xi_2^2}\sqrt{\frac{\lambda_2}{2}}A_0(t_i)\right]^{1/3},
\end{eqnarray}
where $t_i$ are instants when the inflaton $A=0$. Inflaton can decay into $W$ and Higgs bosons only in the vicinity of $A=0$ when the effective masses of these bosons are much less than the inflaton effective mass $\omega$.

$W$ bosons decay into fermions with a decay rate given by:
\begin{equation} \label{Wdecay}
\Gamma_W=0.75 \frac{g^2}{4\pi} m_W,
\end{equation}

while their annihilation cross section is given by:
\begin{equation}\label{Wanni}
\sigma_{WW} \approx \frac{g^4}{16} \frac{2N_l+2N_q N_c}{8\pi \langle m_W^2\rangle} \approx 10\pi \frac{g^4}{16\pi^2\langle m_W^2\rangle}.
\end{equation}

Parametric resonance production of $W$ bosons can start only when their decay rate  in Eq. (\ref{Wdecay}) falls below their production rate through parametric resonance  in Eq. (\ref{Wprod}). Comparing them, we find that resonance production of $W$ bosons can start only when 
\begin{equation}\label{WresC}
A_0<\frac{2}{0.5625\,\pi}\frac{Q^2\,\lambda_2}{\alpha_W^3}A_{cr}\approx 61.88\,\lambda_2\,A_{cr}.
\end{equation}
Production of relativistic particles through decay of $W$ takes place very slowly and would reheat the universe long after the resonance period would have ended \cite{Bezrukov} while production of relativistic particles through annihilation is a much faster process and can yield enough relativistic particles to reheat the universe. Annihilation can occur only when the number density of $W$ bosons is large. This makes the occurrence of parametric resonance necessary allowing us to put a lower bound on $\lambda_2$: 
\begin{equation}
\lambda_2\gtrsim \frac{1}{60}.
\end{equation}

When $W$ is produced through resonance, its number density increases exponentially and the dominant channel for production of fermions is by annihilations of $W$ bosons following Eq. (\ref{Wanni}).

We need to check these conditions for fermion production via Higgs as well. The decay rate of Higgs into fermions is given by the Yukawa couplings:
\begin{equation}\label{hdecay}
\Gamma_h=\frac{y^2}{16\pi}\langle m_h\rangle.
\end{equation}
In Eq. (\ref{hdecay}), only the coupling to top is important as it is large while the coupling for other fermions is very small. The top quark can later decay or annihilate into other fermions. Comparing Eq. (\ref{hdecay}) to resonance production rate of Higgs in Eq. (\ref{hprod}), we find that Higgs production enters the resonance regime only after:
\begin{equation}\label{hresC}
A_0<\frac{64\pi\,Q^2\lambda_2}{\lambda_3\,y^4}A_{cr}\approx0.41\frac{\lambda_2}{\lambda_3}\,A_{cr}.
\end{equation}

Comparing Eq. (\ref{hresC})\footnote{The Eq. (\ref{hresC}) contains only $\lambda_3$ in the denominator instead of full $\lambda_3+\lambda_4/2$ because of our choice of large $\lambda_3$. The fact that Higgs won't be produced via resonance stands even if $\lambda_4$ is used.} to Eq. (\ref{WresC}), we see that if $\lambda_3\lesssim 0.006$, Higgs production will enter the resonance regime around the same time as $W$ boson. For even a small amount of resonance production in Higgs to occur, $\lambda_3$ cannot be greater than $0.41\,\lambda_2$. Since the inert doublet is a dark matter candidate with large couplings, Higgs production will not enter resonance regime till long after the end of the quadratic phase of the potential. The production rate of Higgs remains small and its decay to fermions is at a much lower rate than the annihilation of gauge bosons.

During parametric resonance production of gauge bosons, almost all the $W$ bosons get converted to fermions giving a complete transfer of energy density from $W$ bosons to relativisitc fermions which can be obtained by solving the following equation \cite{Bezrukov}
\begin{equation}
\frac{d(\rho_\gamma a^4)}{dt}=2 a^4 \sqrt{\langle m_W^2 \rangle} \frac{4 Q^2 \omega^2}{\sigma_{WW}},
\end{equation}

which after integration gives
\begin{equation}
\rho_\gamma=\frac{8Q^2\omega^2}{10\pi\alpha_W^2}\frac{6}{13}\left(\frac{4\pi\alpha_W M_{Pl}}{\sqrt{3\lambda_2}}\right)^{3/2}\left[\frac{t_{cr}^{13/6}-t_p^{13/6}}{t_{cr}^{8/3}}\right],
\end{equation}

where $t_p$ is the time when the parametric resonance starts given by the condition in Eq. (\ref{WresC}) and $t_{cr}$ is the end of reheating. During this conversion, almost all the $W$ bosons get converted to fermions so that the only remaining particles by the time reheating ends are the fermions apart from the inert doublet particles. Putting in the numbers, we get:
\begin{equation}
\rho_\gamma\approx \frac{1.46 \times 10^{57}}{\sqrt{\lambda_2}}.
\end{equation}

At this time, energy density in $A$ is:
\begin{equation}
\rho_A=\frac{\omega^2 A_{cr}^2}{2} \approx \frac{1.48 \times 10^{54}}{\sqrt{\lambda_2}}
\end{equation}

We can now obtain the reheating temperature from the energy density in relativistic particles:
\begin{equation}
T_r=\left(\frac{30\,\rho_\gamma}{\pi^2\,g}\right)^{1/4},
\end{equation}
where $g$ is the number of degrees of freedom in the relativistic plasma.

\section{Dark matter}

The end of reheating marks the end of the quadratic oscillations phase of the re-arranged field $A$. Since now, $A$ is the same as $(q^2+x^2)^{1/2}$ and the Jordan and Einstein frames have become equivalent, we can come back to using the physical Jordan frame. The inflaton field no longer has an effective mass $\omega$. Rather things go back to the original inert doublet potential given in Eq. (\ref{OriPot}) with the inert doublet having a mass of $m_2$. In the beginning, the inert doublet obtains a thermal equilibrium with the rest of the relativistic plasma and evolves as radiation. Later, as the temperatures fall and the inert doublet becomes non-relativistic, its evolution is given by the Boltzmann equation. It freezes-out as a cold relic and thus becomes a candidate for cold dark matter.

\begin{figure}[h!]
\centering
\includegraphics[scale=0.4]{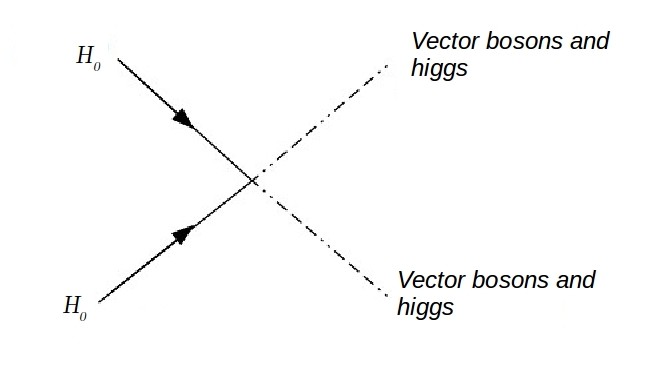}
\caption{\textit{The 4-point vertex interactions}}
\label{Fig4}
\end{figure}

We will use the observed relic abundance of dark matter $\Omega_{dm}\,h^2=0.12$ \cite{Ade:2015xua} to calculate certain parameters in the model. The interactions of the neutral scalar part of the inert doublet are its annihilations into the vector bosons and Higgs. There are no decays of any of the inert doublet components as they are prevented by the $\mathbb{Z}_2$ symmetry. At the tree level, there are 4-point interactions (see Fig. (\ref{Fig4})). The scattering cross-section for these processes is:

\begin{equation}\label{sigv}
\left.\sigma\right|_{CM}v_{rel}=\frac{1}{32\pi\,m_{2}^2\,(1+v_{rel}^2/4)}\sum_{processes}|\mathcal{M}|^2.
\end{equation}

In the non-relativistic limit where $v_{rel}\ll 1$, we can re-write Eq. (\ref{sigv}) as
\begin{equation}\label{SigV1}
\left.\sigma\right|_{CM}v_{rel}=\frac{1}{32\pi\,m_{2}^2}\left(1-\frac{v_{rel}^2}{4}\right)\sum_{processes}|\mathcal{M}|^2.
\end{equation}

The amplitude $\sum_{processes}|\mathcal{M}|^2$ for the neutral scalar component of $\Phi_2$ which is the actual dark matter candidate is:
\begin{equation}
\sum_{processes}|\mathcal{M}|^2=(\lambda_3+\lambda_4+\lambda_5)^2+g^4+(g^2+g^{\prime 2})^2+\frac{1}{8}\frac{g^4g^{\prime 2}}{g^2+g^{\prime 2}}+\frac{1}{2}g^2(g^2+g^{\prime 2}),
\end{equation}

where $g$ and $g'$ are the weak couplings to the vector bosons.

\begin{figure}[h!]
\centering
\includegraphics[scale=0.3]{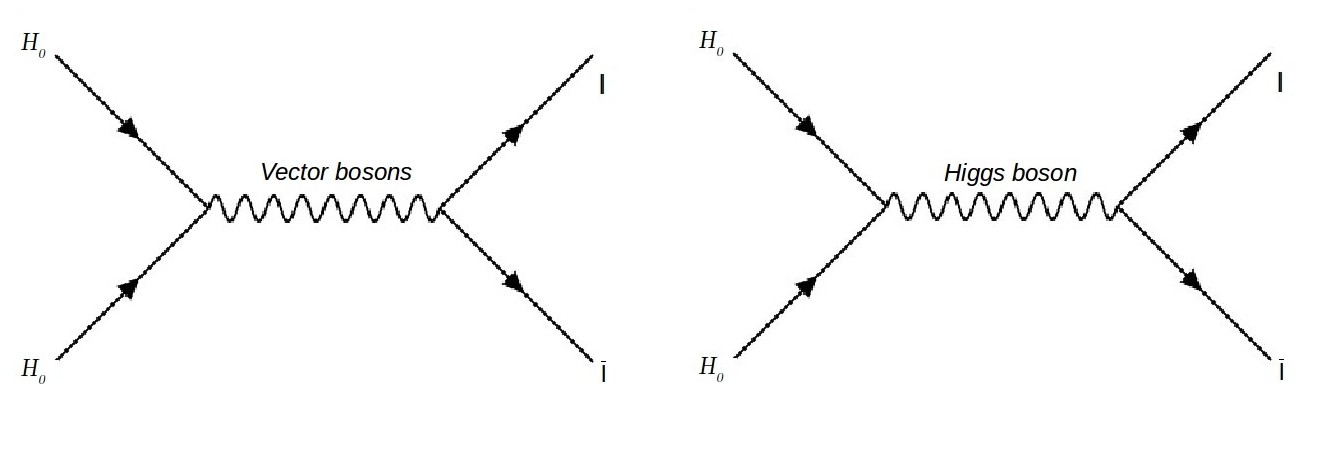}
\caption{\textit{The other tree level interaction diagrams. $H_0$ is the neutral scalar component of the inert doublet.}}
\label{Fig5}
\label{Annihil}
\end{figure}

Apart from these 4-point interactions, there are trilinear couplings as well which include the annihilation of a pair of inert doublet particles via the gauge boson or the Higgs channel into SM particles as shown in Fig. (\ref{Fig5}).

The gauge boson mediated diagrams are momentum dependent. Their contribution is small compared to the one shown in Fig. \ref{Fig4}. Most of the thermal equilibrium evolution of the dark matter particles occur above the EW symmetry breaking scale where the Higgs mediated diagram of Fig. \ref{Fig5} are not present. 

To calculate the freeze-out temperature and the relic abundance, we need to solve the Boltzmann equation. Assuming only $s$-wave annihilations, one can obtain the $x_f=\frac{m_{2}}{T_f}$ \footnote{We use $m_2$ as the mass of the neutral scalar component of $\Phi_2$ because there are no mass corrections which occur only after EW symmetry breaking} at freeze-out where $T_f$ is the freeze-out temperature, and the relic abundance $\Omega_{dm} h^2$ to be \cite{Jungman:1995df,Kolb:1990vq}:
\begin{eqnarray}
x_f&=&\log \left(0.038\, \frac{g}{g_{*s}^{1/2}}\, m_P\, m_{2}\, \langle\sigma v\rangle\right)-\frac{1}{2}\log \left[\log \left(0.038 \,\frac{g}{g_{*s}^{1/2}}\, m_P m_{2} \langle\sigma v\rangle\right)\right],\\
\Omega_{dm}\,h^2&=&1.07\times 10^{9}\,x_f\,\frac{g_{*s}^{1/2}}{g}\,m_P\,\langle \sigma v\rangle,
\end{eqnarray}
where $m_P$ is the Planck mass (not the reduced Planck mass which we have denoted as $M_{Pl}$), $g$ and $g_*$ are the number of degrees of freedom in the plasma and the entropic number of degrees of freedom respectively and $\langle \sigma\,v\rangle$ is taken from Eq. (\ref{SigV1}).

\begin{figure}[h]
\centering
\includegraphics[scale=1.2]{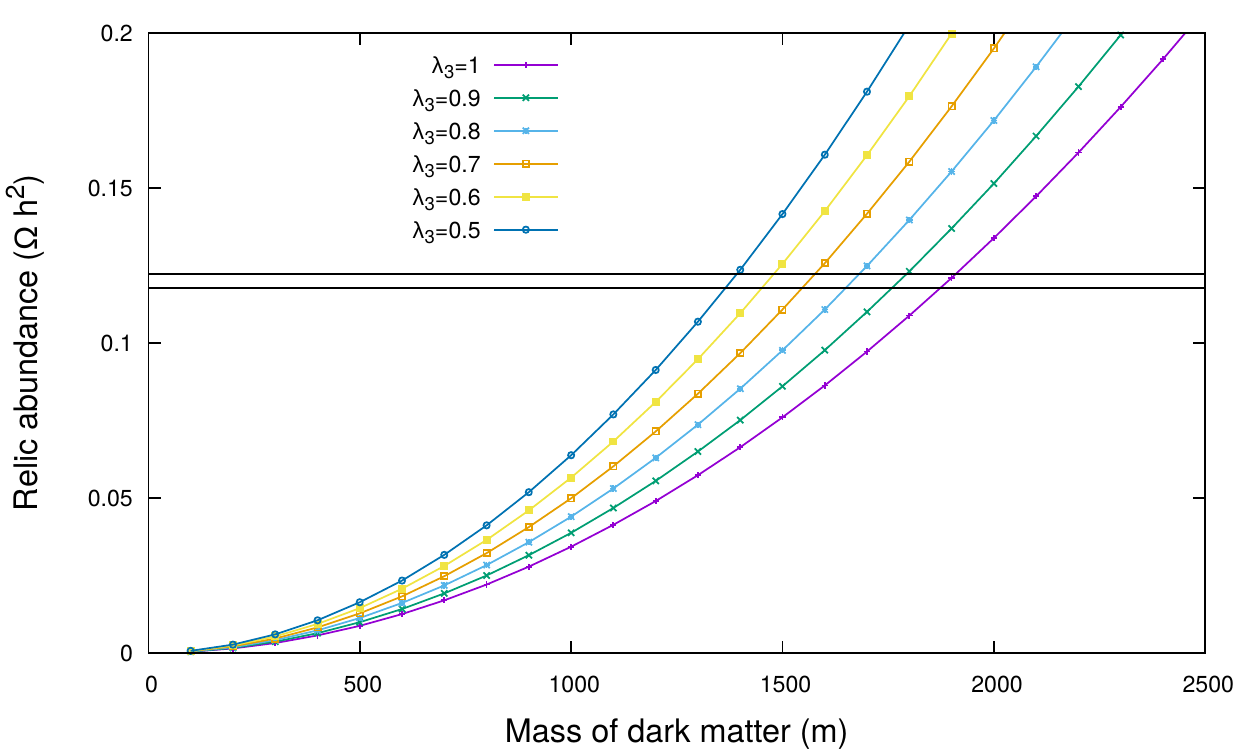}
\caption{\textit{The relic abundance of dark matter vs the mass of the dark matter. The horizontal black band is the Planck 2015 result}}
\label{abundance}
\end{figure}

Planck 2015 data for the relic abundance can now be used to get estimates for the mass of the dark matter and its freeze-out temperature. We obtain:
\begin{eqnarray}
m_{2}=1.89\;\textrm{TeV},\\
T_f=\frac{m_2}{x_f}=65.7\;\textrm{GeV}.
\end{eqnarray}

This calculation has been done using a fixed set of values for $\lambda_3$, $\lambda_4$ and $\lambda_5$ with $\lambda_3\approx 1$ and $\lambda_4,\lambda_5\ll 1$. In principle, $\lambda_3$ can vary between $0.5$ to 1 while still keeping $\lambda_4$ and $\lambda_5$ very small. The effect of varying $\lambda_3$ on the mass of the dark matter candidate is shown in Fig. \ref{abundance} where the solid horizontal line shows the value of the relic abundance obtained from Planck 2015 \cite{Ade:2015xua} and is equal to 0.12. Note that $m_2$ is the dark matter mass till EW symmetry breaks which occurs around the same time as freeze-out. After the symmetry breaks, dark matter mass will get a small correction of order $100$ GeV. The Table \ref{vary} gives the values of dark matter mass satisfying the relic abundance constraint for various values of $\lambda_3$. The corresponding freeze-out temperatures are a little below the EW symmetry breaking scale suggesting that we include the Higgs mediated diagrams in Fig. \ref{Fig5} in the calculations. However, their contribution to the calculations are very small and any changes that they bring about in dark matter masses are beyond the second decimal place.

\begin{table}[h!]
\begin{center}
\begin{tabular}{||c|c||}
\hline 
\rule[-1ex]{0pt}{2.5ex} $\lambda_3$ & Mass in TeV \\ 
\hline \hline
\rule[-1ex]{0pt}{2.5ex} 1 & 1.89 \\ 
\hline 
\rule[-1ex]{0pt}{2.5ex} 0.9 & 1.78 \\ 
\hline 
\rule[-1ex]{0pt}{2.5ex} 0.8 & 1.67 \\ 
\hline 
\rule[-1ex]{0pt}{2.5ex} 0.7 & 1.56 \\ 
\hline 
\rule[-1ex]{0pt}{2.5ex} 0.6 & 1.47 \\ 
\hline 
\rule[-1ex]{0pt}{2.5ex} 0.5 & 1.38 \\ 
\hline 
\end{tabular}
\caption{Effect of varying $\lambda_3$ on mass of dark matter}
\label{vary}
\end{center}
\end{table}

\section{Conclusion}

Explaining inflation and dark matter remain two challenges for any theory beyond the standard model of particle physics. The inert doublet model has been studied extensively in the literature in the context of generating neutrino masses and mixing as well as dark matter. The doublet is called inert because of a $Z_2$ charge assignment which forbids all Yukawa couplings of this doublet with the standard model fermions. This is done to avoid all undesirable flavor violations in the model. In this work we showed that the inert doublet coupled non-minimally to gravity could act both as the inflaton driving slow-roll inflation as well as the cold dark matter of the universe. We obtained a Starobinsky like potential from the model and showed that both slow-roll parameters $\epsilon,\eta\ll 1$. We calculated the scalar power spectrum, the tensor to scalar ratio  and the spectral index in our model and showed them to be well within the observed limits from Planck. After successfully reheating the universe, the inert doublet attains thermal equilibrium and eventually freezes-out as a cold dark matter. We obtained bounds on the couplings of the scalar potential from reheating and dark matter constraints and showed that the Planck bound on relic abundance can be satisfied for neutral scalar component mass of the inert doublet  of around 1.3 to 2 TeV.

\vskip 1cm
{\bf Acknowledgements}\\
The authors would like to thank the Department of Atomic Energy
(DAE) Neutrino Project under the XII plan of Harish-Chandra
Research Institute.
This project has received funding from the European Union's Horizon
2020 research and innovation programme InvisiblesPlus RISE
under the Marie Sklodowska-Curie
grant  agreement  No  690575. This  project  has
received  funding  from  the  European
Union's Horizon  2020  research  and  innovation
programme  Elusives  ITN  under  the 
Marie  Sklodowska-
Curie grant agreement No 674896.

\bibliography{refs}

\end{document}